# Restoring the Broken Covenant Between Compilers and Deep Learning Accelerators

Sean Kinzer
University of California San Diego
skinzer@ucsd.edu

Soroush Ghodrati
University of California San Diego
soghodra@ucsd.edu

Rohan Mahapatra
University of California San Diego
rmahapat@ucsd.edu

Byung Hoon Ahn
University of California San Diego
bhahn@ucsd.edu

Edwin Mascarenhas
University of California San Diego
emascare@ucsd.edu

Xiaolong Li
Qualcomm Technologies, Inc.
lixiao@qti.qualcomm.com

Janarbek Matai
Qualcomm Technologies, Inc.
jmatai@qti.qualcomm.com

Liang Zhang
Qualcomm Technologies, Inc.
liangz@qti.qualcomm.com

Hadi Esmaeilzadeh
University of California San Diego
hadi@eng.ucsd.edu

## 1 Introduction

Deep Learning has taken the IT industry by a storm and it is set to penetrate various disciplines and markets from healthcare [1] and social networking [2] to gaming [3] and entertainment [4]. However, its success is predicated upon the availability of responsive execution platforms as DNNs require massive computations [5, 6]. In fact, they have become the driving use-case for the development and adoption of domain-specific accelerators [7, 8]. These new architectures require state-of-the-art and highly optimized compilers prior to even delivering the expected performance and efficiency gains.

Four challenges make compilers for these designs different than ones targeting conventional general-purpose processors. First, these architectures no longer adhere [9–11] to the long-held abstraction of fine-grained Instruction-Set Architectures (ISAs) and Von Neumann model [12]. Therefore, more micro-architectural features and components need to be exposed, considered, and controlled by the compiler. For instance, an accelerator compute block typically exposes coarser-grained operations than an ALU that performs an individual addition instruction (e.g., a systolic array performs a whole matrix operation). Second, the on-chip storage is no longer a limited set of registers backed by a hardware-managed cache, it is usually several software-managed scratch pads with various access semantics. Third, the interconnection for on-chip data movement and off-chip loads/stores needs to be handled explicitly by compiler, with the appropriate granularity (e.g., tile size). Finally, the compiler needs to match the rather coarse-grained operations (layers) of a DNN to the varying granularity of computation and storage, supported by the hardware.

To address these challenges, one option is to take a software-centric approach [13, 14] by restricting architectures to a standardized ISA that makes the compiler reusable. However, this approach limits the architectural innovations, offering orders of magnitude benefits through novel, specialized execution semantics. Another option is to take a hardware-centric approach [10, 15] that demands re-implementing new compiler stacks and optimization infrastructure for each accelerator.

Alternatively, this paper takes on these challenges and sets out to simultaneously enable the reuse of the compiler while reducing constraints on the architecture. To achieve these conflicting objectives, we propose a compilation framework that integrates a novel architecture abstraction, dubbed the Architecture Covenant Graph (ACG), in its workflow. Traditional ISAs focus on what fine-grained instructions an architecture can perform, which typically operate with a register file and an opaque caching system. In contrast, ACG is defined to capture accelerator structure as a graph consisting of compute units, on-chip/off-chip memory components, and interconnect; each of which contains operational capabilities as attributes.

To leverage this abstraction, we also devise the Codelets construct which is combined with the ACG to enable our Covenant compiler to target varying types of DNN accelerators. While the ACG abstracts the architecture, the Codelets represent the DNN operations and are gradually transformed into accelerator execution schedules by the Covenant compiler. Each Codelet represents DNN layers as sequences of operation on input variables to produce output variables. During compilation, Codelets are transformed into schedules by mapping operands to ACG memory locations, and assigning operations to ACG compute nodes capable of execution. Once operands and operations are mapped to ACG nodes, the dependence between operations and their operands is translated to explicit data transfer operations over the ACG interconnect.

While a number of inspiring works have achieved multi-target compilation and scheduling support [14–16], the requirements for efficiently scheduling and generating code for new targets can be prohibitive. For scheduling to new targets, frameworks such as TVM [14] use flexible, target-agnostic scheduling directives to optimize DNN kernels, but each DNN operator schedule requires hand-tuning by architectural experts. As an alternative to manually scheduling, FlexTensor [17] and then Ansor [18] proposed novel search algorithms capable of identifying optimal schedules using stochastic search and performance measurements, but are inflexible to scheduling on new and unique architectures. Our approach provides the opportunity to adapt these scheduling techniques to new targets and further prune the space of transformations by coalescing architectural characteristics into the schedule. For code generation, both TVM as well as Glow [15] intentionally exclude architectural details because they rely on LLVM [13] as a backend, which is not designed for accelerators. Instead, we provide a malleable technique for code generation which is particularly important for architectures ordinarily using intrinsics which cause powerful instructions to be treated as black boxes by compilers. To support additional accelerators as compilation targets, these frameworks require creation of custom compiler backends and hand-tuned schedule templates.

*Our Covenant compiler is intended for an orthogonal purpose: **automatically** scheduling and generating code for accelerators without a unified, LLVM-like backend by integrating an architecture abstraction into the compiler. This is one of the main contributions of the work, in addition to the ACG and Codelet constructs which enable Covenant to target varying deep learning accelerators.*

To demonstrate the flexibility of the Covenant compiler, we implement ACGs for Qualcomm® Hexagon™ Vector eXtensions (HVX) [1] [19] and an open-source DNN accelerator [10]. For both architectures, we compile 14 different DNN layers across a combination of transformer networks, neural recommender systems, and convolutional neural networks and measure their performance. When targeting HVX, our automated approach achieves 93.8% of the performance of TVM's hand-scheduled templates that rely on manually constructed intrinsic. Compared to manually-implemented DNN layers in Qualcomm Technologies' nnlib which include hand-written assembly kernels, we achieve 31.3% improved performance. Besides HVX, we target an open-source DNN accelerator [10] that shows the flexibility of the Covenant compiler to target an entirely different architecture. The Covenant compiler achieves 182× performance improvement using the DNN accelerator compared to a CPU baseline. Finally, we illustrate the feasibility of implementing optimizations using the Covenant compiler by combining different optimization passes and achieve 128.6× speedup compared to unoptimized code on HVX. These results show the flexibility of the Covenant compiler for automating scheduling and code generation for accelerators while maintaining high-performance by integrating architecture characteristics through the ACG and Codelets.

## 2 The Missing Link: An Abstraction for Micro-Architecture Specification

General-purpose processors are based on the von Neumann model of computing, which is a sequential fine-grained instruction execution model. Hence, compilation for these processors is made possible by exposing the Instruction Set Architecture (ISA), through which the micro-architecture is completely abstracted away. However, rapidly emerging DNN accelerators tend to use other models of computing, such as systolic in the case of Google's TPU [7, 20] and dataflow in the case of Microsoft's Brainwave [21, 22]. These DNN accelerators typically consist of one or more arrays of Processing Elements, that can only perform simple arithmetic operations in parallel, as shown by the example in Figure 1b. Typically these PEs are connected to one another as well as on-chip memory through software-managed interconnection and memory hierarchy. As such, compilation for these novel architectures requires exposing more of the microarchitectural details. In contrast, general-purpose processors use a pipeline to enable a number of ALUs to carry out instructions, as illustrated in Figure 1a. They are also connected to the memory through a hardware-managed cache. The fundamental differences in the compute model and the organization of the architecture and microarchitecture between DNN accelerators and general-purpose processor clearly demonstrates the need for a new abstraction for compilation. However, exposing every detail makes compiler design an adhoc practice for each specific microarchitecture that is not reusable. Instead, DNN accelerator abstractions are required to enable a reusable compilation

[1] Snapdragon and Qualcomm branded products are products of Qualcomm Technologies, Inc. and/or its subsidiaries.

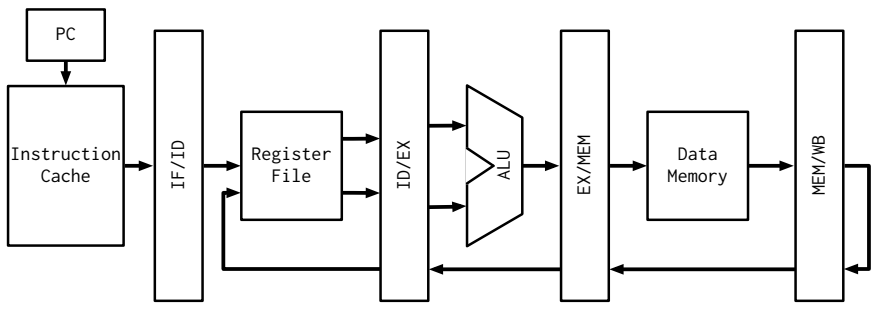

**(a)** Pipeline processor microarchitecture.

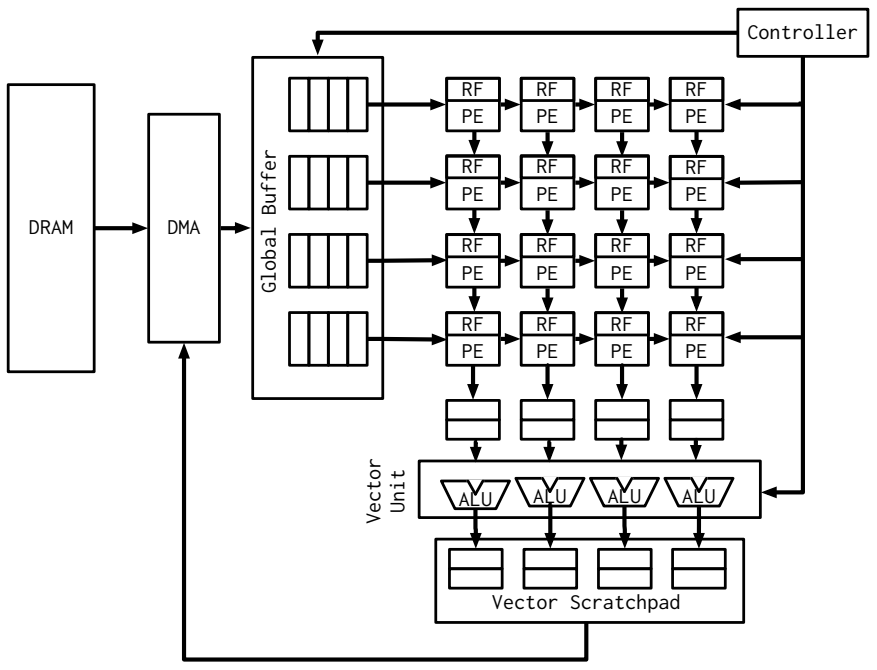

**(b)** Example DNN accelerator microarchitecture.

**Figure 1.** Comparison of microarchitectures for general purpose processors and DNN accelerators.

workflow for different types of DNN accelerator microarchitecture. The following section details such an abstraction, called the Architecture Covenant Graph (ACG).

### 2.1 Architecture Covenant Graphs

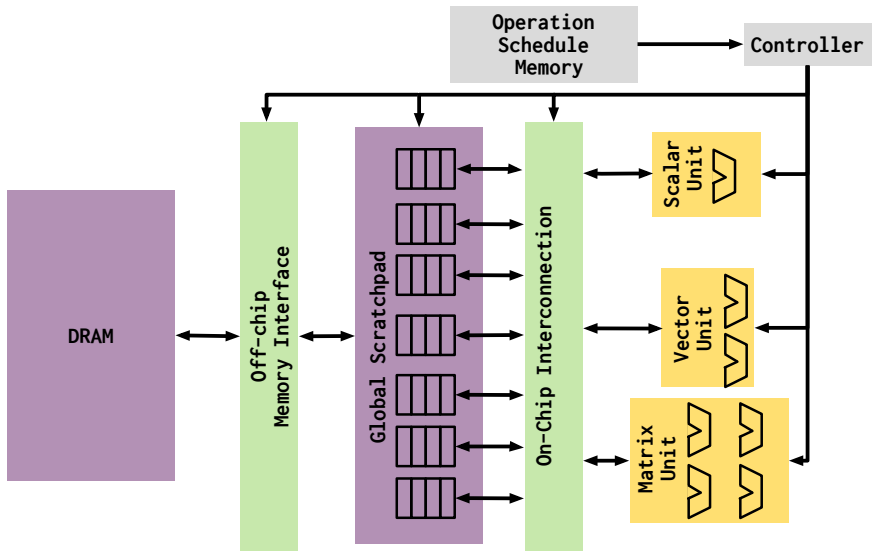

**(a)** Example DNN accelerator architecture.

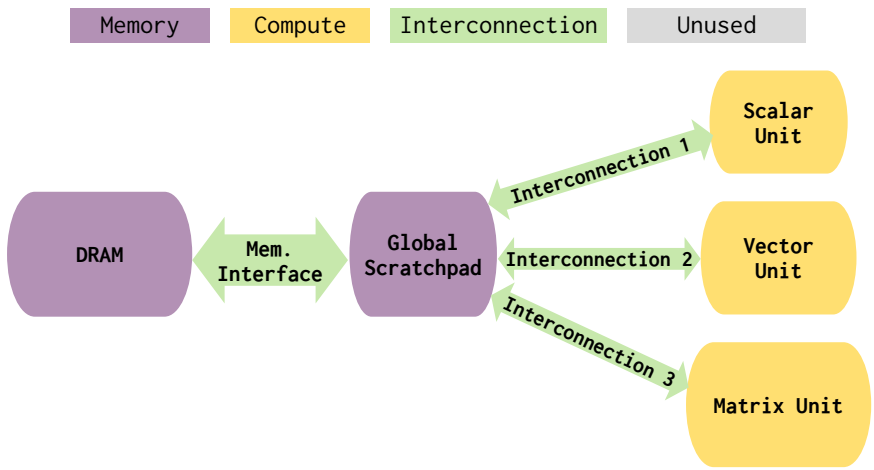

**(b)** Architecture Covenant Graph (ACG) example.

**Figure 2.** Example DNN accelerator architecture and its ACG.

We describe ACG and it's design rationale by using a running example of a generic DNN accelerator microarchitecture and the corresponding ACG in Figure 2. Figure 2a visualizes the microarchitecture for an example DNN accelerator, including its off-chip memory and software-managed, on-chip memory in purple, programmable interconnection in green, and three functional units with unique capabilities in yellow.

To capture the data movement properties on DNN accelerator microarchitectures such as these with programmable interconnection and different types of functional unit for mapping operations to, the ACG is modeled as directed graph as shown in Figure 2b. Each ACG is comprised of vertices representing *programmable memory* and compute components, and unidirectional or bidirectional edges connecting each component. The edges represent the programmable interconnection between the on-chip/off-chip memory and compute components. Edge direction is required for enabling a reusable compilation workflow, as it informs the scheduler of valid paths for moving data, such as DRAM to `Global Scratchpad`, and `Global Scratchpad` to one of the functional units in Figure 2a. In this example, for each of the interconnections, data can be read and written to and from each of the functional units (`Scalar Unit`, `Vector Unit`, `Matrix Unit`) and the `Global Scratchpad`, as well as between DRAM and the `Global Scratchpad`. In some other cases multiple on-chip scratchpads are used for different purposes, with some scratchpads being restricted to sending data to functional units and unable to receive data, in which case the edge between them would be unidirectional. This is unlike traditional memory and caches in general-purpose processors, which are passive and generally do not execute instructions to send or receive data. Instead, the processor core is the active party that loads or stores data to these passive structures. In contrast, the compiler for a DNN accelerator often needs to generate instructions for memory components since they are active elements. Figure 2a also includes three separate programmable functional units capable of executing separate operations in parallel: a `Matrix Unit`, a `Vector Unit`, and a `Scalar Unit`. By using a directed graph, the compiler is capable of identifying opportunities for parallelizing operations across multiple functional units by selecting graph nodes which support the operation and have a common memory node predecessor.

However, scheduling the data movement also requires validation that the size of data being transferred is able to fit on the intermediate storage nodes such as `Global Scratchpad` in Figure 2a because there is no hardware-controlled data caching mechanisms. To distinguish between the attributes necessary for computation versus memory, ACG uses compute nodes shown in yellow and memory nodes shown in purple, each of which have distinct sets of attributes for informing the compilation process. In addition, lower-level architecture components shown in gray in Figure 2a such as the `Controller` for sending control signals to other components and `Operation Schedule Memory` for storing operations are not included in the ACG. *With the primary goal being machine code generation, the ACG excludes components such as these and other low-level details because they are not programmable, and do not provide relevant information to the compiler.*

Lastly, the unique properties across different DNN accelerator microarchitectures and even across their functional units binds them closely to the binary codes necessary for execution. As an example, the `Matrix Unit` in Figure 2a uses dataflow execution to perform matrix multiplication, only requiring data availability from the scratchpad to execute instead of relying on an explicit matrix multiplication binary code. In addition, making data available may require a sequence of binary codes for separately sending each input data to the functional unit rather than a single, dedicated code. Therefore, the ACG specifies binary code for a DNN accelerator as mnemonics without tying them to a specific computation model or set of execution semantics. This allows the code generation implementation to be reused across different architectures by because sequences of mnemonics can be defined for a finite set of operations which are delineated by the ACG nodes and edges.

Below, the specification used for mnemonics is detailed, in addition to the different attributes of compute nodes, memory nodes, and edges, included in the ACG.

#### 2.1.1 Memory.

Software-controlled memory such as `Global Scratchpad` in Figure 3 allows the compiler greater control over data reuse, but also require explicit mnemonics for operations such as off-chip data transfers. To ensure valid memory accesses during execution, the access semantics and capacity of the memory needs to be known to the compiler so that memory request addresses are properly aligned. As shown in Figure 3, each memory node includes attributes defining their access semantics, such as the `data_width` for specifying the smallest unit of accessible data in bits is 32. The `data_width` is particularly important for DNN accelerators supporting mixed precision operations, because certain functional units might support 16-bit operations but read data from a memory component storing each 16-bit operand with a 32-bit `data_width`. In this case, the compiler must ensure that 16-bit operands are stored in 32-bit chunks rather than packed together, and the increased memory consumption for the operation is accounted for.

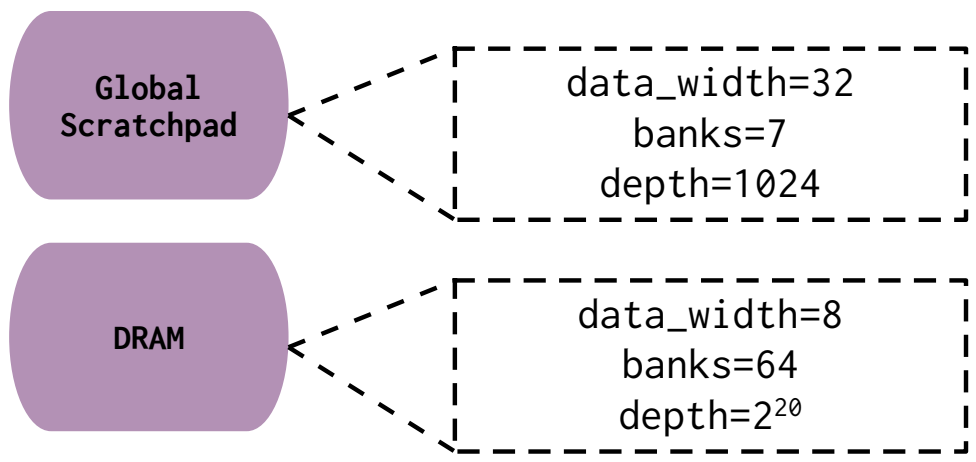

**Figure 3.** ACG storage nodes and their capabilities.

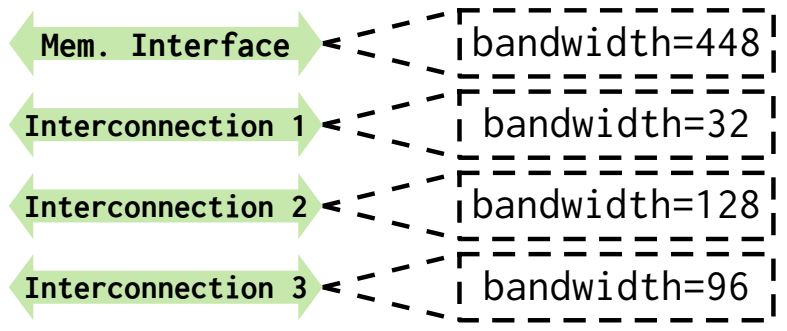

**Figure 4.** ACG interconnection examples

In addition, memory nodes use the `banks` attribute to denote the number of banks in a memory component, as it is common for on-chip memory to include varying number of banks for reading and writing multiple data in parallel to/from coarse grained functional units such as the `Vector Unit` or `Matrix Unit` shown in Figure 2b. Each bank is capable of sending `data_width` bits of data at a time, which means `data_width` × `banks` determines the size of an addressable element in the memory component. When selecting the sizes of on-chip data to be stored and operated on, the compiler must use this information to ensure the size is correctly aligned in memory by requiring data chunks are divisible by the size of an addressable element. As an example, the `Global Scratchpad` has $32 \times 7 = 224$ bit entries, which must be taken into account when generating mnemonics requiring address calculation based on immediate values.

Finally, compilers can exploit large on-chip scratchpads for data reuse by partitioning operands into chunks called tiles which are stored on-chip and operated on together. To validate tile selection, the compiler must ensure all being stored at once is within the capacity of the on-chip memory being used. For the `Global Scratchpad`, the capacity can be calculated by multiplying the `depth` attribute by the addressable element size: $224 \times 1024 = 229{,}376$ bits, or 28,672 bytes.

**2.1.2 Interconnection.** When it comes to generating code for transferring data on and off a DNN accelerator, a single binary code is often insufficient due to the limitations imposed by the interconnection between on and off-chip memory. For instance, `DRAM` in Figure 2a is connected to `Global Scratchpad` through a bidirectional `Off-Chip Memory Interface` interconnection. This link constrains the amount of data in bits transferred at a time, or may allow for more than one unit of `Global Scratchpad` data to be moved in a given cycle. In the running example, a directed edge called `Mem. Interface` represents these types of interconnection which represent the supported programmable communication capabilities. The directed ACG edges use the `bandwidth` attribute to define the amount of data in bits capable of being transmitted in a single operation as shown in Figure 4. This information is crucial during compilation, as DNN accelerators provide more flexible data transfer capabilities allowing variable-sized data transfers between on and off-chip memory. Furthermore, the bandwidth determines the number of memory requests the compiler needs to generate for this specific edge to load a tile of data.

In addition, the `Interconnection` is capable of sending data to multiple parallel programmable functional units, with unique data processing properties, therefore requiring different bandwidths. To distinguish between the different data transmission properties between a single interconnection and different DNN accelerator components, the ACG includes several `Interconnection` edges with unique bandwidths.

This is particularly important when making scheduling decisions, because a coarse-grained operation could be mapped to multiple parallel functional units with hardware-controlled synchronization, but the interconnection between on-chip storage and certain functional units may require multiple data transfer operations for sending the necessary operand data.

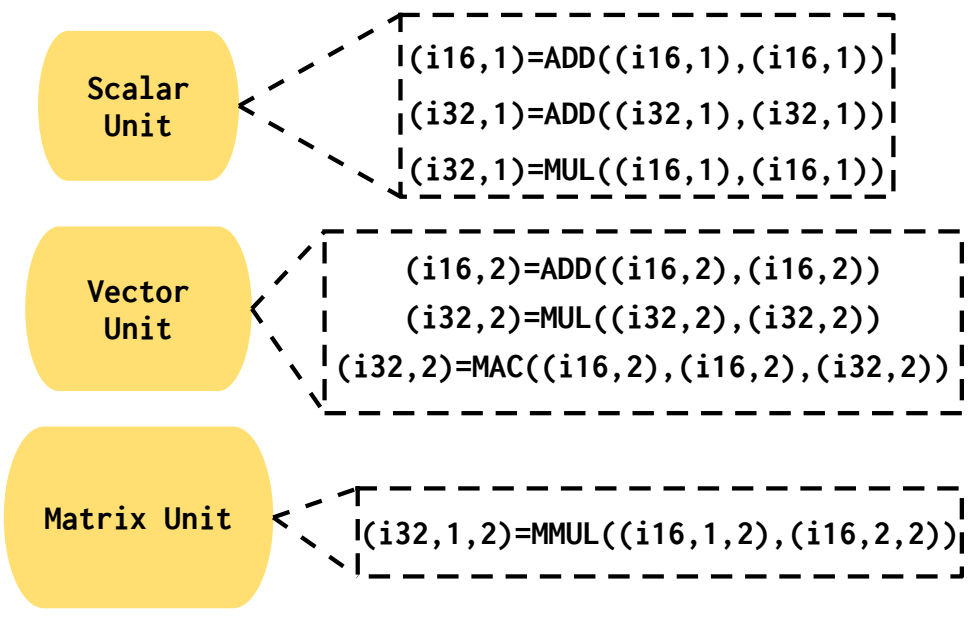

**Figure 5.** ACG compute nodes and their capabilities.

**2.1.3 Compute.** DNN accelerators provide unique opportunities for mapping coarse-grained operations to a variety of compute resources, as shown in Figure 2a, which includes a 2×2 `Matrix Unit`, 2-wide `Vector Unit`, and `Scalar Unit`. The ACG represents programmable functional units as compute nodes, using an attribute called `capabilities` to describe the coarse-grained functionality supported by the corresponding architecture component. Figure 5 demonstrates capabilities for each compute in Figure 2b, with each compute node supporting varying granularity, datatype, and number of operations. Capabilities encapsulate opportunities for parallelism and type-specific operations in the compute nodes. They are defined by an operation name and an ordered list of datatype and element size pairs for each input/output operand associated with the operation. A subset of the supported operations are defined in Table 1. For example, the `Vector Unit` supports the `ADD` operation, taking two input operands with two, 16-bit integer elements and generates two 16-bit integer output elements. The sizes and datatypes are included in the operand specification because the specialized compute units in DNN accelerators are capable of performing different operations in parallel on varying kinds of operand datatypes and sizes. By defining capabilities this way, the compiler can identify which functional units can execute parts of the DNN layer in parallel by matching the operation name and data type to the functional unit capability, and then breaking the coarse grained DNN operation into the same size chunks. To demonstrate this, consider an element-wise addition operation specified as: `(i16,3)=ADD((i16,3),(i16,3))`. The compiler can decompose this operation into a scalar addition on the `Scalar Unit` and a vector addition on the `Vector Unit`, as both compute nodes support 16-bit integer addition at different granularities. To ensure the full range of layer mappings are exposed to the compiler, capabilities defined for a compute do not require one-to-one mappings between capability primitive and a functional unit's mnemonic. As an example, the `Vector Unit` might not directly support a multiply-accumulate (`MAC`) operation using a single mnemonic, but it can be defined as a capability by breaking it into separate multiply-add mnemonics.

**Table 1.** Subset of supported capabilities and their definitions.

| Type | Name | Description |
|---|---|---|
| Unary | RELU | Rectified Linear Unit function. |
| | SIGMOID | Logistic sigmoid. |
| | TANH | Hyperbolic tangent function. |
| Binary | ADD/SUB | Element-wise addition and subtraction. |
| | MUL/DIV | Element-wise multiplication and division. |
| | MAX/MIN | Element-wise maximum/minimum. |
| | MMUL | Matrix-matrix multiplication. |
| Ternary | MAC | Multiply-accumulate. |
| | GEMM | General Matrix Multiply. |

**2.1.4 Mnemonics.** Thus far, the ACG has described the structure and programmability of a DNN accelerator, but the mnemonics which can be composed to carry out the data movement and operations represented in the ACG must also be defined to generate executable binaries. In contrast to general-purpose processors which use instructions and assume a von Neumann compute model, different DNN accelerators depend on different compute models with unique machine code semantics. Thus, machine codes for a DNN accelerator are defined as mnemonics stored as an ACG attribute for generating sequences of mnemonic code. Each individual mnemonic is defined with customizeable attributes for analysis/optimization, and an ordered list of named fields with fixed bitwidth, which can represent either a constant number or an enumerated set of values. As an example, a mnemonic with the `ADD` id is defined above and includes 4 fields, where `src1`, `src2` and `dst` are constant fields representing the starting addresses in scratchpad, and `target` is an enumerated value field which can be set to one of `SCALAR` or `VECTOR` depending on the functional unit to be executed on. By generically defining mnemonics in this manner, they can be used for different types of DNN accelerators without binding the mnemonics to certain execution semantics.

$$mnemonic \in Mnemonic ::= \textbf{mnemonic}\ name(opcode)\{field^{*}, attr^{*}\}$$

**(a)** Mnemonic definition syntax.

```
# Example: ADD #3,#0,#1, VECTOR
mnemonic ADD(3) {
  ifield("SRC1_ADDR",8),
  ifield("SRC2_ADDR",8),
  ifield("DST_ADDR",8),
  efield("TGT", 1, ["SCALAR","VECTOR"])
}
```

**(b)** Example of `ADD` mnemonic definition.

**Figure 6.** Example of a mnemonic definition.

# 3 Codelets

To flexibly enable DNN compilation to domain-specific architectures, a programming abstraction must capture both the semantics of an operation, and the relevant microarchitecture components it is tied to. In addition, a construct for enumerating the different types of macro-mnemonics required for code generation must be designed. Covenant uses compute kernel abstractions called Codelets which are complimentary to the ACG to enable compilation. Codelets are defined prior to compilation as a sequence of operations on parametric-shaped operands called *surrogates* which represent DNN layers. Initially, the operations do not include architecture-specific details, which enables their portability across different architectures. However, during Covenant compilation each Codelet is gradually transformed to define the sequence and mapping of operations based on an ACG. Codelets are declared using a DNN layer name, and are composed of `compute`, transfer, and loop operations which represent operations on tensors, movement of data, and repetition of operations. As an example, an add Codelet can be defined as shown in Figure 7a. To integrate ACG information into the compiler, Codelet operations rely on different types of surrogate variables to encompass both data attributes (e.g., datatype, shape) and ACG location throughout execution.

```
cdlt add {
  N=param();
  a=inp([N],null,null);
  b=inp([N],null,null);
  c=out([N],null,null);
  loop n(N) {
    c[n]=compute(null,"ADD",a[n],b[n]);
  }
}
```

**(a)** Initial Codelet.

```
cdlt add {
  # Size and datatype are set
  a=inp([12],i16,null);
  b=inp([12],i16,null);
  c=out([12],i16,null);
  loop n(12) { # Number of iterations is set
    c[n]=compute(null,"ADD",a[n],b[n]);
  }
}
```

**(b)** Codelet mapped to a DNN layer.

**Figure 7.** Example of a add Codelet.

### 3.1 Surrogate Variables

The process for generating valid sequences and mappings of operations on data is inherently tied to accelerator attributes. As such, surrogate variables in Codelets encode shape information, datatype and ACG location:

```
x=inp([dim1,...,dimN],dtype,loc);
```

Data movement is tracked by requiring surrogate variables to be associated with a single ACG location, defined by the loc attribute. Using single location surrogates has the added benefit of distinguishing between a DNN layer input and a tile generated from operand data because they will each be represented by different variables with a similarly different shape and layout in memory. To further simplify Codelet compilation, different types of surrogate variable with unique semantics are used. For instance, Codelets are defined with the basic assumption that it will receive a certain number of inputs and generate outputs, defined as inp and out type surrogates. In addition, each DNN layer performs similar sets of operations on different shaped tensors, and sometimes use parameters to define how the layer is executed, both of which are represented as param surrogates. Prior to compilation, other unset fields such as the location and datatype are set to null to indicate they have not been assigned. When a relu layer is mapped to a Codelet, each param surrogate is replaced with with the corresponding layer-specific value, which results in known input/output sizes and operation sizes as shown in Figure 7b The datatype is also set during the layer mapping, and is assumed to be provided in the DNN layer specification.

Once a Codelet has been mapped to a DNN layer instance, the Covenant compiler applies ACG attributes to transform the Codelet. To begin, the compiler assumes that inp and out surrogates are stored on the highest level of the ACG memory hierarchy, identified as the memory node with the longest path to each functional unit. Once the operand surrogates are mapped, computations are mapped to compute nodes in the ACG and data movement operation to and from the target compute node are added. The last surrogate type, locals, represent data stored on the intermediate nodes on the path from an inp location to compute node location or compute node location to an out location. Each local surrogate is created as a result of transfer and compute operations, and their attributes are inferred based on the source operation as we will discuss below.

### 3.2 Codelet Operations

DNN accelerators provide diverse sets of programmable compute and memory resources for enabling more parallelism when it comes to computation and additional opportunity for data reuse due to programmable memory. In contrast to von Neumann architectures which use uniform memory accesses and sequential computation models, scheduling for DNN accelerators adds an extra layer of difficulty by having to keep track of where data is stored and where computation is occurring. Codelets address these complexities using three categories of operations to represent DNN operations: loop, transfer and compute. Each operation operations type has a fixed set of attributes which determine layer-specific and architecture-specific properties required for scheduling and code generation. To demonstrate how the Covenant compiler uses these constructs to handle the added complexity we will reference the example shown in Figure 8. In the example, the add Codelet shown in Figure 8b is targeting the ACG in Figure 8a, and the the starting location of inp and out surrogates is set to MEM1.

***Compute operations:*** To accommodate the variation in operations supported by different compute units, the compute operations are defined using the coarse-grained capabilities described in Section 2.1.3, and operate on tensor operands. Every compute operation has is defined with the ACG compute node it is mapped to, it's capability name, and the surrogate operands and their offsets:

```
c[i]=compute(loc,capability, op1[off1], op2[off2],...,opN[off3]);
```

To specify tensor offsets, compute operands can be indexed using loop operations, which can be converted to address offsets for programmable memory by combining the size of the surrogate and the addressing information for it's ACG location.

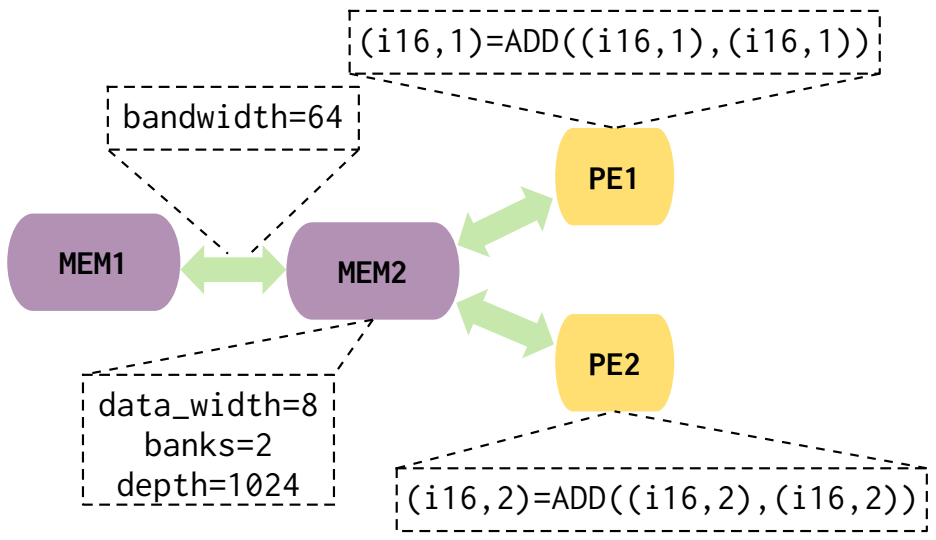

**(a)** Example ACG use for compilation.

```
cdlt add {
  a=inp([12],i16,"MEM1");
  b=inp([12],i16,"MEM1");
  c=out([12],i16,"MEM1");
  c1=transfer(i16(0),"MEM2",[2]);
  loop n(0,6,2) {
   # Using only 25% of bandwidth!
   a1=transfer(a[n],"MEM2",[2]);
    # Much more available memory!
   b1=transfer(b[n],"MEM2",[2]);
   # Unused PE!
   c1[0]=compute("PE2","ADD",a1,b1);
   transfer(c1,c[n],[2]);
  }
}
```

**(b)** Codelet prior to being fully scheduled.

```
cdlt add {
  a=inp([12],i16,"MEM1");
  b=inp([12],i16,"MEM1");
  c=out([12],i16,"MEM1");
  c1=transfer(i16(0),"MEM2",[6]);
  loop n(2,stride=6) {
    # Tile loops, maximum bandwidth use
   a1=transfer(a[n],"MEM2",[6]);
   b1=transfer(b[n],"MEM2",[6]);
   loop n1(2,stride=3) {
     # Split
       operations between PE1 and PE2, in parallel
     c1[n1]=compute("PE2","ADD",a1[n1],b1[n1]);
     c1[n1+2]=compute("PE1","ADD",a1[n1+2],b1[n1+2]);
   }
    transfer(c1,c[n],[6]);
  }
}
```

**(c)** Scheduled Codelet

**Figure 8.** Example Codelet scheduling using an ACG.

The ADD compute operation in Figure 8b can be mapped to either PE1 or PE2, as both include the supported capability but with different granularities. The compiler automatically determines the mapping by selecting the ACG node capable of performing the most operations at a time, PE2 in this case, because it can do two element-wise additions at a time. Once selected, the Covenant compiler updates the location field in the compute operation with the target compute node.

***Transfer operations*** After mapping each compute operations in the Codelet, the compiler orchestrates data movement across programmable memory by adding explicit transfer operation to the Codelet. transfer operations are used in Codelets to represent data movements across a DNN accelerator, explicitly codifying scheduling of data locations as required by domain-specific compilers. In Figure 8b, this can be accomplished by first finding the shortest path between MEM1 and PE2 and adding transfer operations for each operand and each edge. transfer operations are specified with a source, destination, and the transfer size in number of source elements in each dimension of the source operation. The semantics of a transfer operation can differ depending on the type of source and destination used, which accommodates the different operations required by programmable memory:

```
dst=transfer(src[i],"MEM1", [n]); # Move data to MEM1
dst=transfer(i16(0),"MEM1", [n]); # Allocate new memory at MEM1
transfer(src[i],dst[i], [n]); # Overwrite data stored in dst
```

In the first two examples, the destination is specified by an ACG node name, which tells the compiler that memory needs to be allocated at that location and a new local surrogate is generated. When the source of allocation transfers is an operand with an index offset, the compiler will generate a local surrogate with n elements as its size, the same datatype as the src, at MEM1. Alternatively, new memory can be allocated for reuse if the source operand is a constant value which includes its type and size. In cases where the destination is an operand such as dst, memory at dst location will be overwritten and no additional surrogates will be generated.

To determine the needed transfer type for Figure 8b, the Covenant compiler adds transfer operations for each of the edges between the source operand location and compute target. Specifically, the compiler generates new memory allocation operations necessary for storing the outputs of ADD on MEM1 as shown in Figure 8c. The compiler also generates transfer operations for each inp to the intermediate memory nodes. Lastly, the compiler must send the on-chip results stored in MEM2 to MEM1, which will write data to the location of c. Once inserted, each transfer operation can be combined to calculate the cumulative size of the data for each memory node at different points in the Codelet by tracking the transfer operations and their sizes, as well as operand datatypes.

***Loop operations:*** Lastly, to represent repeated operations, addressing offsets, and operations execution amounts, loop operations with the same semantic meaning as "for" loops are used. To satisfy the need for operand address offsets, loop operations can be used as indices for operands in both transfer and compute operations, which is commonly used in DNN operation descriptions to specify which tensor elements are being operated on. Each loop operation is used for specifying DNN layer semantics, and therefore does not include attributes relating to the ACG. loop operations are created using a variable name, lower and upper iteration bounds, the stride, and opening a scope for execution using curly braces:

```
loop i(0,6,2) { ... } # Iterate from 0 to 6, stride=2
```

To use loops as indices for surrogates, loop name is put inside brackets alongside a surrogate to represent an address offset.

A key component of compilation is tiling, and `loops` offer a familiar construct to apply tiling transformations, as loop splitting is a commonly technique for tiling on general purpose processors. When tiling a Codelet, `loops` are split into groups according to the number of transfers required to send data from it's source to the compute destination. Splitting a `loop` operations consists of factoring the number of iterations into an outer `loop` operations with a step size corresponding to how large a tile will be, and an inner `loop` operations which has a range equivalent to the outer loops step size.

### 3.3 Macro-Mnemonics

For DNN accelerators, generating valid mnemonics is conditioned on which functional unit is being used because the same operation can generate different mnemonics depending on the compute unit. The Covenant compiler ensures valid code generation by combining operations types, operand types, and their ACG node attributes to select pre-defined functions for generating sequences of mnemonics called macro-mnemonics. These macro-mnemonics use the Codelet operation type it is matched with, the ACG node(s) it is associated with, and the containing Codelet as contextual input to define mnemonics generation. Each mnemonic is generated by populating it's fields with either statically determined values or by using attributes of Codelet operations.

## 4 Enabling Optimization

Compiler optimizations for DNNs have been shown to enable significant performance improvements when targeting CPU and GPU [18, 23]. However, state-of-the-art, stochastic optimization techniques which rely on performance measurements to guide the algorithm cannot be applied to domain-specific architectures without the ability to generate executable code. When targeting domain-specific architectures, optimizations have the potential to offer even greater benefits due to their tendency to provide more compute and memory resources with greater programmability.

The Covenant compiler is intended to be a community driven project which improves as a crowd-sourced effort. Therefore, the initial goal is to provide a framework which *enables* new and existing optimization algorithms to be constructed and benefit from the use of the ACG rather than introducing new optimizations. Below, we discuss how existing optimizations can be transformed by integrating architectural details into the algorithm.

Codelet optimization passes are defined as functions which take an individual Codelet and the ACG as arguments, and return the transformed Codelet. Providing the ACG as an argument allows for retrieval of certain characteristics embedded in the ACG because Codelet operations only contain the ACG node names as attributes. The attributes embedded in ACG nodes bolster common optimizations used in traditional compilers which might otherwise be applied using a heuristic.

**Algorithm 1:** Codelet Tiling Validation

```
function ValidTiling(codelet, ACG)
    let V ← ∅ // Valid tilings
    let f_i = loop iteration factors for loop_i ∈ codelet
    let P = factor permutations ∈ f_i
    for each p ∈ P do
        let constraint_sat = True
        // Keep track of data stored on each ACG storage node
        let storage[s] = 0  for each storage node s ∈ ACG
        for each t = transfer ∈ codelet do
            let p_t = {factor ∈ p | factor ∈ t.offsets}
            let xfer_size = t.operand.dtype.bits × Π(p_t)
            storage[t.dst] += xfer_size // Update t.dst storage
            if (xfer_size mod t.src.data_width) ≠ 0 or
              storage[t.dst] > t.dst.capacity then
                constraint_sat = False
                break
        if constraint_sat == True then
            V = V + p
    return V
```

***Tiling Validation*** is one example of commonly used optimizations is loop tiling, where loops are grouped into smaller blocks of operations on tiles of data to increase the data locality as previously discussed. In contrast to other frameworks, tiling is built into Covenant scheduling algorithm rather than being an optimization pass, although further optimization of tile selection can be implemented as a Codelet optimization. Here, we show how tiling validation is performed in the Covenant compiler, and can be extended to search-based optimization passes. When tiling operations for targeting general purpose processors, loop ranges can be split using almost any permutation of numbers which are factors of the loop iterations because memory is typically hardware-controlled which prevents invalid memory requests. In contrast, domain-specific architectures often provide programmable memory where certain tiling permutations will lead to invalid programs instead of slower programs. As shown in Figure 1, Covenant validates tiling by first collecting all valid factors of the Codelet `loop` ranges in $f_i$, and then generating all unique combinations of those factors in $P$. The first concern for tile validation is sufficient memory space to store each tile, which is a map of memory ACG nodes to data sizes,$storage$, is initialized to 0 to track total storage for the permutation $p$. Each factor in $p$ represents a possible stride for a `loop` operation, and each `transfer` operation uses `loop` operation as index offsets. This allows the `transfer` size to be computed using the product of `loop` strides and the datatype size of the `transfer` operand, *t.operand.dtype.bits*. Once the $transfer_size$ is computed, it is added to the destination memory node in $t$. The tiling can be validated by first checking if the $transfer_size$ is divisible by the data width of the source memory node to ensure addressability, and then the capacity of destination memory node is verified. If these constraints are satisfied, the tiling permutation is validated and is added to a set of possible tilings for final scheduling.

***Loop Unrolling*** Loop unrolling is another common optimization, used to reduce the impact of loop branching as well as memory overheads by transferring more data in a loop body and unrolling computations for the transferred data. Using the ACG, opportunities for loop unrolling can be identified by iterating over `transfer` operations, and checking the bandwidth of the edge connecting the source and destination ACG nodes. If the `transfer` size is less than the edge bandwidth, more data can be transferred in a single operation if the destination ACG node does not reach maximum capacity.

***Parallelization*** A central focus of domain-specific architectures for DNNs is providing as many opportunities for parallelism as possible. Taking advantage of the parallelism in such architectures is not always trivial, especially when heterogeneous compute cores are available with varying capabilities. However, the ACG simplifies parallelism identification through the capability attributes in compute nodes, which can be combined to form the equivalent operation and therefore be performed in parallel. As an example, Figure 9a

```
cdlt relu {
  a = inp("DRAM",[25],i32);
  c = inp("DRAM", [25], i32);
  loop i(25) {
    c[i]=compute("PE","RELU",a[i]);
  }
}
```

**(a)** Pre-scheduled operations

```
cdlt relu {
  a = inp("DRAM",[25],i32);
  c = out("DRAM", [25], i32);
  loop i(25,stride=5) {
    # SIMD: RELU((i32,4),(i32,4)), PE: RELU((i32,1),(i32,1))
    c[i]=compute("SIMD","RELU",a[i]);
    c[i+4]=compute("PE","RELU",a[i+4]);
  }
}
```

**(b)** Parallelized operations

**Figure 9.** Parallelization Identification Using an ACG.

demonstrates an ReLU operation on two 25-element tensors targeting an ACG composed of two compute nodes: a "SIMD" capable of performing four ReLU operations at a time, and a processing engine ("PE") capable of a scalar ReLU. The two tensors do not factor perfectly into the SIMD, which demonstrates a common difficulty when trying to identify parallelization. One solution to this problem is to introduce additional operations which pad zeros to each of the tensors so that they can be tiled correctly. Instead, the ACG can be used to identify other compute units, namely "PE", capable of being combined with the SIMD to form tiles of parallel operations.

***Mnemonic Packing*** For micro-architectures using Very Long Instruction Words (VLIW), multiple instructions can be performed in parallel by "packing" them together. In these architectures, compiler needs to identify independent instructions and pack them to increase utilization. With Codelet operation being coarsely defined to represent multiple mnemonics, forming mnemonic packets is performed during code generation as an optimization. To form mnemonic packets, ACG resource availability as well as mnemonic dependencies need to be identified. To enable packing, the ACG node executing each mnemonic is identified to determine the resources consumed by a VLIW packet and integrated into the packing algorithm. For dependency analysis, the `field` attributes in mnemonics can be annotated with read and write semantics to identify sequences of independent mnemonics. Using both of these mnemonic attributes allows packet formation by iterating over `mnemonics` for a Codelet and creating a packet with a single mnemonic occupying the `tgt` resource. Then, independent mnemonics capable of execution in the current packet, determined by the consumed ACG resources and available VLIW slots, can be hoisted into the current packet.

# 5 Evaluation

## 5.1 Experimental Setup

***Benchmarks.*** To evaluate covenant, we use a comprehensive set of benchmarks from various classes of DNNs including image classification (InceptionV3 [24], ResNet-50 [25]), object detection (MobileNetV3 [26]), natural language processing (BERT-Large [5]), and neural recommendation systems (DLRM [27]). For image classification and object detection networks we choose convolutional and fully-connected layers that make up the majority of these networks. For BERT-Large, we benchmark the GEMM layers and the self-attention layer of an encoder block. Finally, for DLRM, we benchmark its Multi-Layer Perceptron (MLP) fully-connected layers. Table 2 lists all the DNN layer benchmarks with their layer dimensions. N shows the sequence length for language models and the batch size for other DNNs. IW/IH and OW/OH show the input/output width/height dimensions of the layers, while KW/KH parameters specify the weight kernel dimensions. Note that for FC/GEMM layers, these dimensions are equal to one. Finally, IC/OC column show the number of input/output channels for the DNN layers. We use INT8 precision for inputs/weights and INT32 precision for outputs of layers.

### 5.1.1 Target Architectures.

To demonstrate the flexibility of Covenant for multi-target compilation, we use two distinct architectures: HVX [19] and an open-source DNN accelerator [10]. For each architecture, we use the ACG DSL for Covenant compilation.

***DNNWeaver.*** DNNWeaver is a parameterizable DNN architecture which consists of two main compute components: (1) a systolic array connected to several on-chip buffers that is capable of executing various-sized convolution and GEMM layers, and (2) a SIMD vector processing array connected to two vector scratchpad memories that supports the remainder of layers (e.g. pooling, activation, normalization, etc.) As shown in

**Table 2.** DNN Layer Benchmarks.

| Model | Layer | N | IH/ IW | OH/ OW | KH/ KW | IC/ OC | # Heads |
|---|---|---|---|---|---|---|---|
| BERT-LG | GEMM1 | 384 | 1 | 1 | 1 | 1024/4096 | - |
| | GEMM2 | 384 | 1 | 1 | 1 | 4096/1024 | - |
| | ATN1-GEMM | 384 | 1 | 1 | 1 | 1024/64 | 16 |
| | ATN2-GEMM | 384 | 1 | 1 | 1 | 64/384 | 16 |
| | ATN3-GEMM | 384 | 1 | 1 | 1 | 384/64 | 16 |
| | ATN4-GEMM | 384 | 1 | 1 | 1 | 1024/1024 | 1 |
| DLRM | FC1 | 1 | 1 | 1 | 1 | 745/367 | - |
| | FC2 | 1 | 1 | 1 | 1 | 367/512 | - |
| | FC3 | 1 | 1 | 1 | 1 | 512/256 | - |
| | FC4 | 1 | 1 | 1 | 1 | 256/1 | - |
| InceptionV3 | FC1 | 1 | 1 | 1 | 1 | 2048/1000 | - |
| | CONV1 | 1 | 299 | 149 | 3 | 3/32 | - |
| MobileNetV3 | CONV1 | 1 | 224 | 112 | 3 | 3/16 | - |
| | CONV2 | 1 | 112 | 112 | 3 | 16/64 | - |
| ResNet50 | FC1 | 1 | 1 | 1 | 1 | 512/1000 | - |
| | CONV1 | 1 | 224 | 112 | 7 | 3/64 | - |
| | CONV2 | 1 | 224 | 56 | 3 | 64/64 | - |

Figure 10a, the systolic array is connected to four separate on-chip buffers by unidirectional edges, where it reads input activation data, model weights, and bias data from IBUF, WBUF, and BBUF buffers, respectively, and writes output to OBUF buffer. Additionally, the SIMD array is connected to OBUF with a unidirectional edge to consume its data, while is also connected with bidirectional edges to two scratchpad memories (VMEM1/2) to read/write vectors during computations. Table 3 lists a subset of attributes for DNNWeaver ACG nodes.

***HVX.*** HVXis a Digital Signal Processor (DSP) created by Qualcomm Technologies, which uses VLIW instructions and includes vector extensions. Figure 10b illustrates the ACG of HVX. As shown, HVX incorporates a scalar core that supports a diverse set of scalar instructions (Add, Mul, MAC, Max, etc.) and uses a General Register File (GRF) for operand read/write. In addition to the scalar core, HVXincludes an additional SIMD processor for vector instructions, with 32 lanes each capable of performing a range of four 8-bit operations to a single 32-bit operation per lane. As opposed to DNNWeaver where all the data transactions between DRAM and on-chip buffers are

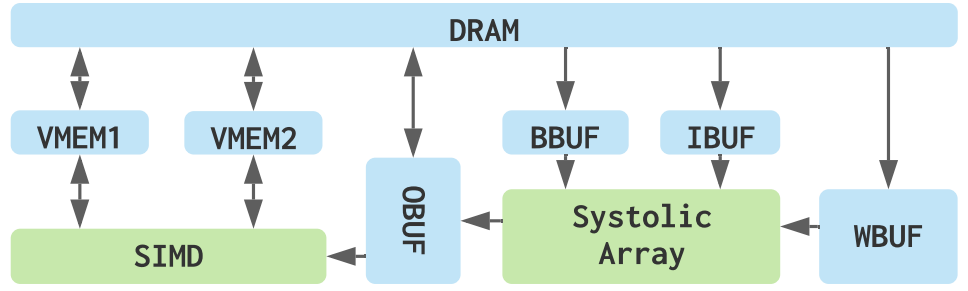

**(a)** DNNWeaver ACG

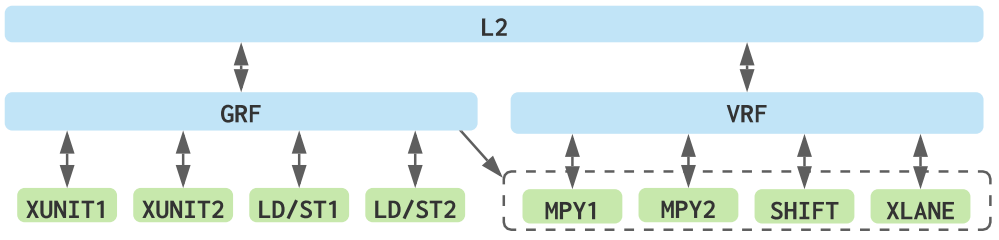

**(b)** HVX ACG.

**Figure 10.** Visualization of ACGs for DNNWeaver and HVX. Blue nodes are memory and green nodes are compute.

**Table 3.** A Subset DNNWeaver and HVXACG Attributes.

| Architecture | ACG Node | Example Attributes |
|---|---|---|
| DNNWeaver | Systolic Array | (i32,64)=GEMM((i8,64),(i8,64,64),(i32,64)) |
| | SIMD | (i32,64)=ADD/SUB((i32,64),(i32,64)) |
| | | (i32,64)=SIGMOID/TANH((i32,64)) |
| | VMEM1/2 | data_width=32; banks=64; depth=2048 |
| | IBUF | data_width=8; banks=64; depth=2048 |
| | WBUF | data_width=8; banks=4096; depth=4096 |
| | OBUF | data_width=32; banks=64; depth=2048 |
| | BBUF | data_width=32; banks=64; depth=1024 |
| | DRAM | data_width=8; banks=1; depth=32 billion |
| Hexagon | CORE | (u8,8)=ADD((u8,8),(u8,8)) |
| | | (i32,1)=ADD((i32,1),(i32,1)) |
| | | (i32,1)=MAC((u8,4),(u8,4),(i32,1)) |
| | | (i32,1)=MUL((i32,1),(i32,1)) |
| | HVX | (i32,32)=ADD/SUB((i32,32),(i32,32)) |
| | | (i32,32)=MVMUL((u8,32,4),(u8,4)) |
| | | (i32,32)=GEMM((u8,32,4),(u8,4),(i32,32)) |
| | | (u32,32)=GEMM((u8,32,4),(u8,4),(u32,32)) |
| | GRF | data_width=32;banks=4;depth=32 |
| | VRF | data_width=1024;banks=32;depth=32 |
| | L2 | data_width=8;banks=32;depth=1024 |

governed explicitly by the instructions, HVXis similar to typical general-purpose processors and incorporates hardware-managed caching mechanisms for loading/storing from/to DRAM, which is why DRAM is not included in the ACG.

### 5.1.2 Performance Measurements and Comparisons.

***Baseline frameworks.*** We compare the performance of our proposed Covenant compilation framework to two other frameworks: nnlib and TVM. For all three comparison points we use the HVXas the target architecture and evaluate the performance of the compiled benchmark DNN layers. For benchmark baselines, we use optimized PyTorch [28] implementations on an Intel Xeon E7 CPU. nnlib [29] is a framework developed by Qualcomm Technologies for offloading DNN operations to HVX, comprising a set of hand-tuned C code and assembly kernels for DNN layers. TVM [14] is a compilation stack that supports a variety of general-purpose architectures as well as its own custom accelerator, VTA [30]. To compile to HVXusing TVM, we used hand-tuned schedules and manually defined intrinsics developed by Qualcomm Technologies' experts, which generate optimized LLVM code for HVX.

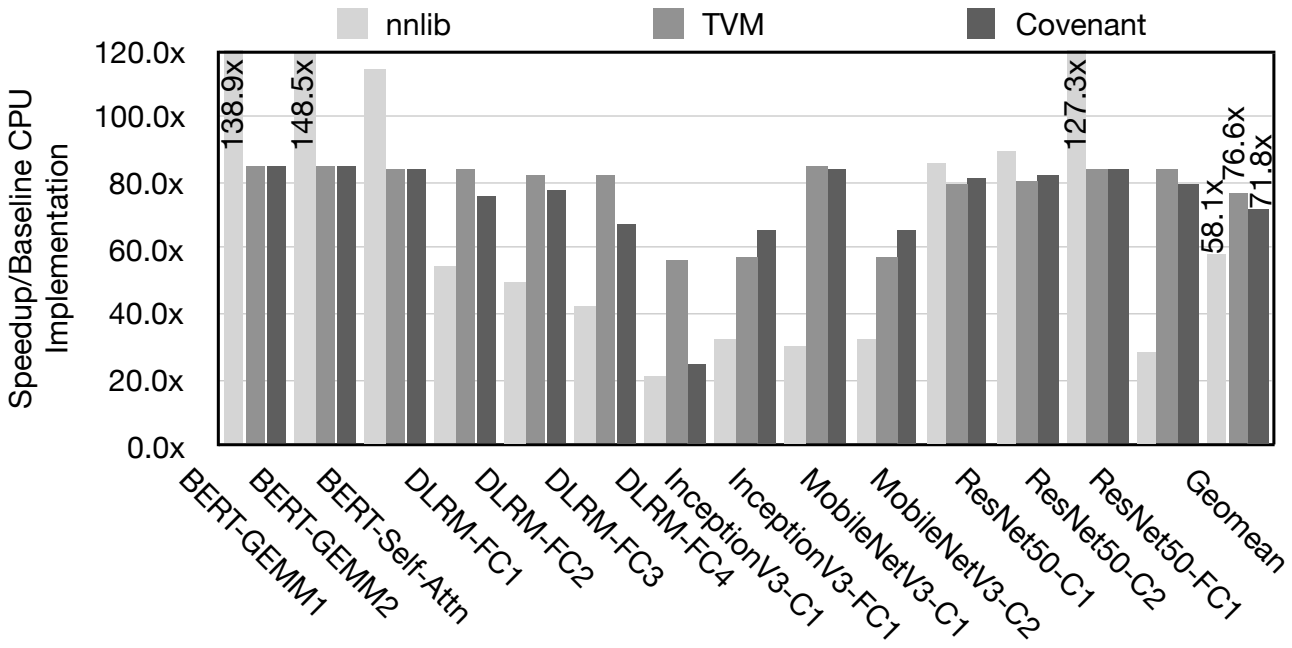

**Figure 11.** Performance comparison of various frameworks.

***Performance measurements.*** To measure the performance of the codes compiled by Covenant, nnlib, and TVM targeting HVX, we use the built-in cycle-accurate Hexagon SDK simulator developed by Qualcomm Technologies' experts. To assure a fair comparison, we include the device execution time, which manifests the actual runtime of the DNN layers on the target hardware, for all the comparison points and use that without considering host execution overheads. To evaluate the capability of the Covenant framework in targeting multiple architectures, we also use DNNWeaver, an open-source DNN accelerator [10]. To measure the runtime of the Covenant compiled code on DNNWeaver, we used its open-sourced cycle-accurate simulator [31]. To verify the correctness of the compiled codes for all the frameworks and target architectures, we compare the outputs generated by the simulators with the software implementation of the DNN layers in PyTorch.

## 5.2 Results

**5.2.1 Framework Comparison.** Figure 11 shows the speedup enabled by the three compilation frameworks targeting Hexagon DSP, compared to a baseline CPU implementation. Across all benchmarks, Covenant provides an average of 31.3% improvement compared to nnlib's hand-tuned kernels. Covenant also achieves 93.8% of TVM's performance on average. As Figure 11 shows, all three frameworks perform better on larger layers having more operations. This results from the compounding parallelization optimizations across more loop iterations. Among all benchmarks, BERT-GEMM1 and BERT-GEMM2 layers see the maximum performance gains, as the larger number of computations in these layers provide highest code optimization opportunities. Relative to TVM, the DLRM-FC4 has a smaller speedup in Covenant because it includes a branch instruction for the single-iteration OC loop, whereas TVM generated code avoids this overhead. With regard to nnlib, the improvements are more significant for larger layers, due to inclusion of hand-tuned tensor transformations allowing more MAC operations per cycle. However, these transformations can be detrimental for smaller layers (e.g., DLRM) and convolutional layers where the total size of reduction dimensions is smaller because the transformations cannot maximally utilized, and the overhead is magnified. Lastly, TVM is also able to achieve consistent speedups across each benchmark, similar to Covenant, with the added advantage of LLVM optimization passes. As a result, TVM manages to achieve high performance for even small benchmarks such as DLRM-FC4, but does not attain the significant speedups of nnlib which required specialized tensor transformations.

**5.2.2 Optimization Results.** We evaluate the effectiveness of three Codelet optimizations when targeting HVX [19]. Figure 12 shows the benefits across the benchmark DNNs enabled by the optimizations. The baseline is vanilla Covenant implementations for the DNN layers. We first use Vectorization based on the parallelization techniques described in

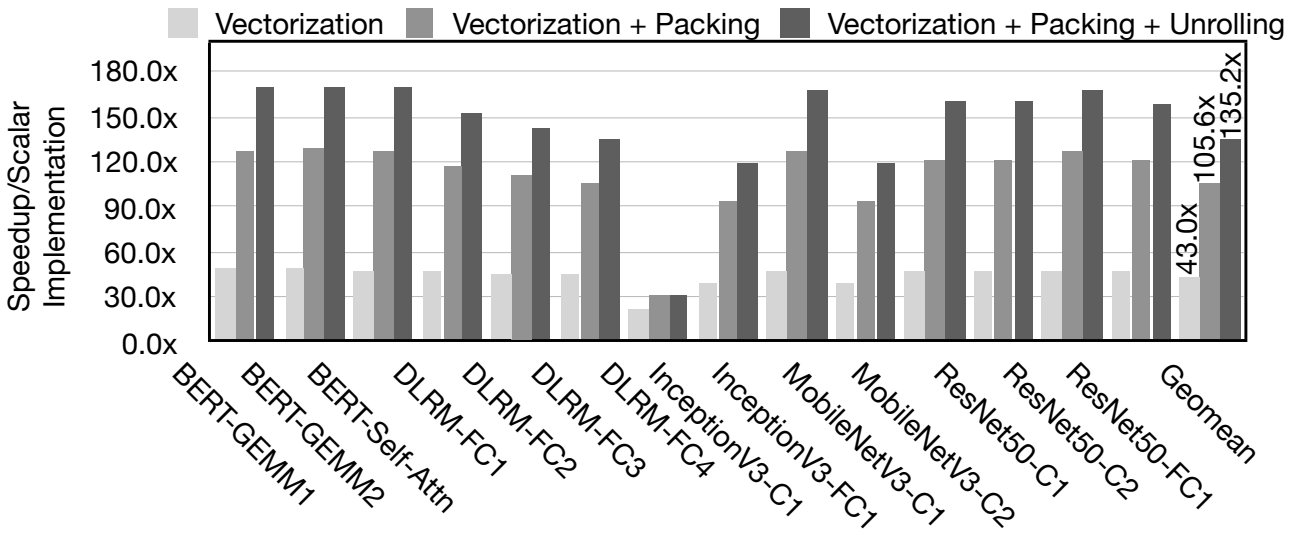

**Figure 12.** Performance improvements based on code optimizations implemented in Covenant.

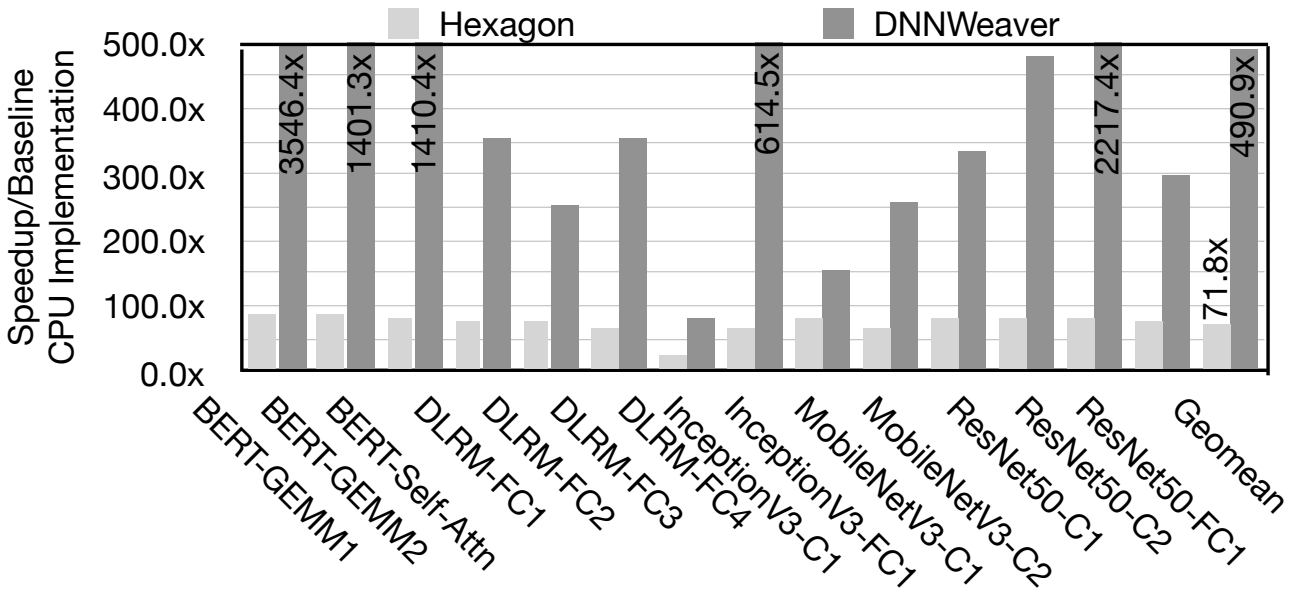

**Figure 13.** Performance results of evaluated hardware, while using Covenant for compilation.

Section 4. We then enable Mnemonic Packing, as described in Section 4, on top of Vectorization. Finally, we add the third optimization, Loop Unrolling as discussed in Section 4. As the figure 12 shows, Vectorization is the most effective technique. This is a due to massive data-level parallelism available in both DNN layers, as well as HVX. Among the benchmarks, DLRM-FC4 sees the least improvement due to its relatively smaller matrix dimensions. On average across all DNN layer benchmarks, Vectorization achieves 43.0× speedup compared to the baseline CPU implementation. Mnemonic Packing leverages the mnemonic level parallelism opportunities in compiled DNN mnemonics to utilize the four available instruction slots in HVXarchitecture. On average, it brings about an additional 2.4× performance improvements. Finally, Loop Unrolling is enabled to facilitate efficient memory accesses, which provides a 1.3× extra performance improvements, on average.

**5.2.3 Multi-Target Compilation.** To demonstrate the flexibility in targeting various hardware architectures, we use Covenant to compile to two different styles of architectures. Hexagon DSP is a more general-purpose-style architecture that supports a wide range of operations. On the other hand, DNNWeaver is a domain-specific specialized DNN accelerator with a systolic array architecture that only supports DNN execution. Figure 13 shows the performance of these two hardwares compared to the baseline CPU implementation. On average, HVXbrings 71.8× speedup over baseline CPU, while DNNWeaver provides 490.9× performance improvements, both using Covenant for compilation. The higher speedups offered by DNNWeaver are due to two reasons: 1) DNNWeaver harbors 32× more number of compute resources compared to HVXand 2) it utilizes a systolic array architecture which is specialized for vector-matrix multiplications, as opposed to SIMD architecture of HVX. Across all benchmarks, DNNWeaver performance improvements are more pronounced for larger DNN layers, as they require large matrix multiplications, suitable for systolic array architectures.

## 6 Related Work

With the growing interest in DNN accelerators, creating efficient and flexible compilers for them is increasingly vital. This work fundamentally differs from prior works in that it integrates a novel accelerator architecture abstraction (ACG) into the compilation stack through Codelets construct. These two enables seamless reuse of the same compiler across various accelerators. Below, we discuss the most related works.

***Compiler Infrastructure for DNN Accelerators.*** MLIR [32] and Glow [15] seek to enable compilation for different targets by offering multiple levels of IR. However, they fall short of code generation due to not offering a mechanism to describe the target hardware. Tensorflow's XLA [16] is another framework that uses a high-level graph IR for compilation to general-purpose processors and domain-specific Google's TPUs. Similarly, XLA is a set of optimizations on a specialized IR that is a representation of the DNN and does not concern itself with abstractions for the hardware (i.e., ACG and Codelets).

***Architecture Abstractions for Scheduling.*** A prior work has leveraged architecture abstractions for scheduling on spatial architectures by modeling them as directed graphs [33]. This work is focused solely on scheduling methodology and does not deal with code generation, whereas Covenant comes with a complete compilation stack that leverages Codelets to facilitate use of scheduling techniques for code generation.

***Architecture Abstractions for Hardware Generation.*** A number of prior works have used DSLs to incorporate architecture features into algorithm specification for the purposes of hardware generation [34–36]. LLHD [37] uses MLIR [32] to simplify hardware design and generation by defining an architecture description language. Covenant fundamentally differs from these prior works because it aims to leverage architecture abstractions to compile to various existing hardware as opposed to generating new hardware.

***Low-level IRs for DNN Scheduling.*** Halide [38] and it's extensions [39] introduced the idea of distinguishing between computation and schedule to compile image processing pipelines, and include schedule transformations for common optimizations. TVM [14] takes inspiration from Halide and uses tensor expressions combined with additional scheduling operations such as tensorization to optimize code generation. Schedules for tensor expressions in TVM's IR support arbitrary transformations regardless of the target backend, but can be constrained with manual construction of a valid schedule templates for each tensor expression to constrain code generation [40]. Tensor Comprehensions [41] and PlaidML [42] automate the scheduling process using tensor-based IRs, yet lack flexibility for scheduling to new hardware. These works do not propose or integrate an architecture abstraction into the compiler. Moreover, in contrast to these IRs, the Covenant compiler performs scheduling by integrating architectural details into Codelets, enabling scheduling algorithms be reused across DNN operations and different targets.

***Schedulers for DNN Operations.*** Another body of works have focused solely on scheduling for different architectures. FlexTensor [17] and Ansor [18] automate the scheduling process by extending TVM's code generation backend. However, they cannot perform scheduling for the accelerators without a pre-existing compiler and runtime environment. Fireiron [43] is a scheduling language for targeting to only GPUs that explicitly incorporates data movement into schedule definitions. CoSA [44] is a scheduling framework that incorporates hardware features into a mixed-integer programming algorithm to form constraints on schedules, without support for code generation. In contrast, Covenant compiler leverages the combination of ACGs and Codelets to provide a uniform and automated compilation framework with code generation backend for targeting to various DNN accelerators.

## 7 Conclusion

DNN accelerators are introducing a new age of compiler design requiring alternative constructs and abstractions. This paper defines two such building blocks, ACG and Codelets. The ACG is an architecture abstraction which makes various components of the accelerator and their connectivity accessible to the compiler. The ACG is integrated into the Covenant compiler through the Codelet construct which represents mutable operations on DNNs, and is progressively transformed into execution mappings and schedules on the ACG. The encouraging empirical results show this work is an effective step towards developing compilers, targeting different accelerators.